\documentclass{appolb}

\usepackage{amsfonts,amsmath,amssymb,bm}
\usepackage{graphicx}

\begin{document}
\title{MINIMAL SPANNING TREE GRAPHS AND POWER LIKE SCALING IN FOREX NETWORKS}
\author{A. Z. G\'orski$^a$, S. Dro\.zd\.z$^{a,b}$, J. Kwapie\'n$^a$,
P. O\'swi\c ecimka$^a$
\address{ %
$^1$H. Niewodnicza\'nski Institute of Nuclear Physics, Polish Academy of Sciences, Radzikowskiego 152, Krak\'ow, 31-342, Poland \\%
$^2$Institute of Physics, Univ. of Rzesz\'ow, 35-310 Rzesz\'ow, Poland}}
\maketitle
\begin{abstract}
Correlation matrices of foreign exchange rate time series are investigated for 60 world currencies.
Minimal Spanning Tree (MST) graphs for the gold, silver and platinum are presented.
Inverse power like scaling is discussed for these graphs as well as 
for four distinct currency groups (major, liquid, less liquid and non-tradable).
The worst scaling has been found for USD and related currencies. 

\end{abstract}



 The foreign currency exchange (FOREX, FX) market is the world largest
financial market and the exchange rates have direct influence on all other
markets because the price of any asset is expressed in terms of a currency.
The FX market is strongly decentralized (transactions
take place in many different places in the world), commonly accessible and
extremely difficult to control. In addition, there
is no friction (transactions are basically commission free).
Due to time differences FX transactions are performed 24 hours a day,
5.5 day a week with maximum between 1 and 4 p.m. GMT, when both American
and European markets are open. Hence, the FX time series are especially
worth of detailed analysis.
 The FX market can be viewed as a complex network of mutually
interacting nodes, each node being an exchange rate of two
currencies and being strongly interconnected with complex
nonlinear interactions to other nodes.
Any currency can be expressed in terms of a given
currency that is called the {\em base currency}.

For a financial time series of an i$th$ asset ($i=1, \ldots, n$)
at time $t$, $x_i(t_i)=x_i$, one defines its return over the time period
$\tau$ as
$G_i(t;\tau) = \ln x_i(t+\tau) - \ln x_i(t)$,
where the logarithm is used to have additivity with respect to the
return time $\tau$.
For the FX series instead of a value $x_i(t)$ one has $x^B_A(t)$,
an exchange rate, {\em i.e.} a value of currency $A$ expressed in terms
of a base currency $B$ and instead of $x_i(t)$ one has $x_A^B(t)$.
The returns can be denoted as $G_A^B(t)$ and they are clearly
antisymmetric: $G_A^B(t;\tau) = - G_B^A(t;\tau)$ and they fulfill the
{\it triangle rule}:
$G_A^B(t;\tau) + G_B^C(t;\tau) + G_C^A(t;\tau) = 0$
\cite{APP2006}, already for relatively small values of $\tau$.
As a result, for a set of $n$
currencies we have $N=n-1$ independent values and the same number of
nodes with a given base currency.

 We study time series of daily data for 60
currencies, including gold, silver and platinum \cite{DataSource}.
The data were taken for the time period Dec 1998--May 2005 and the
series were filtered to get rid of misprints.
In particular, we have removed daily jumps greater than $5\sigma$ (less than
0.3\% of data points). Also, the gaps related to non-trading days were
synchronized. For each exchange rate we have obtained a time series of
1657 data points.
The currencies are denoted according to ISO 4217 standard, and they can be
divided into four groups, according to their liquidity.
The major currencies, we call the $A^\star$ group, include USD, EUR,
JPY, GBP, CHF, CAD, AUD, NZD, SEK, NOK, DKK (11 currencies).
To the group $A$ belong all other liquid currencies: CYP, CZK, HKD, HUF,
IDR, ILS, ISK, KRW, MXN, MYR, PHP, PLN, SGD, SKK, THB, TRY, TWD, XAG,
XAU, XPT, ZAR (21 currencies). Less liquid currencies (group $B$)
include: ARS, BGN, BRL, CLP, KWD, RON, RUB, SAR, TTD (9 currencies).
Finally, the non-tradable currencies (group $C$) taken into account
are: AED, COP, DZD, EGP, FJD, GHC, HNL, INR, JMD, JOD, LBP, LKR, MAD,
PEN, PKR, SDD, TND, VEB, ZMK (19 currencies).
Exchange rates of these currencies can be viewed as a complex network
of mutually interacting nodes.


The (symmetric) correlation matrix can also be computed in terms of the
normalized returns, $g_A(t)$. To this end one has to form $N$ time
series
$\{ g_A^X(t_0), g_A^X(t_0+\tau), \ldots, g_A^X(t_0+(T-1)\tau)\}$
of length $T$. These series can form an $N\times T$ rectangular matrix
$M$. The CM can then be rewritten in the matrix notation as
\begin{equation}
\label{CM2}
{\bm C}^X \equiv  [C]_{AB}^X = \frac{1}{T} \ {\bm M}^X {\widetilde{\bm M}^X} \ ,
\end{equation}
where tilde stands for the matrix transposition.
By construction the trace of the CM equals the number of time series
$\text{Tr} \ {\bm C}^X = N$.

\begin{figure}
\begin{center}
  \includegraphics[width=11.0cm,angle=0]{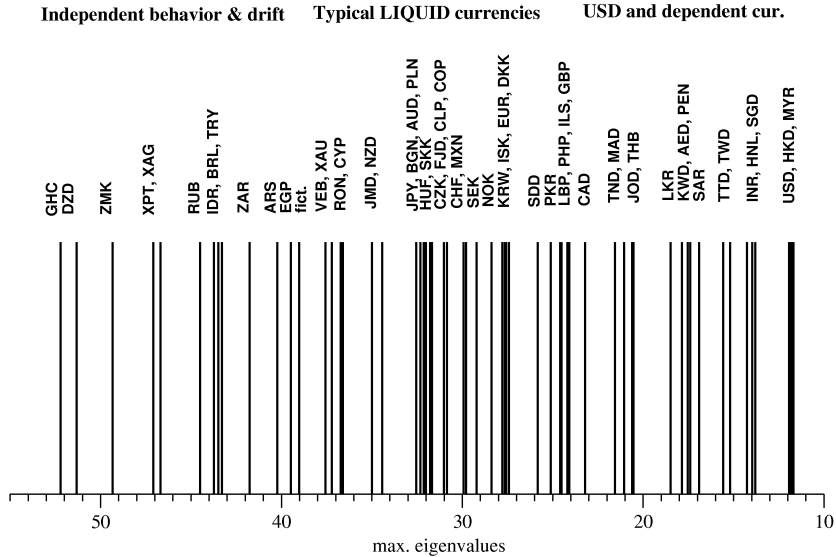}
  \caption{Maximal eigenvalues of CM for all base currencies, including the fictitious (Gaussian noise) currency. Three types of behavior are indicated: independent drift, typical behavior and USD-tied currencies.}
\end{center}
\label{fig:fig1}
\end{figure}

 Maximal eigenvalues for all base currencies considered are plotted in
Fig~1. The plot can be divided into three parts. The largest
maximal eigenvalues correspond to the currencies that either have a very strong drift,
independent of the behavior of other currencies (like GHC) or whose
fluctuations are to large extend independent of the global FX behavior.
In the medium range (approximately $\lambda_N = 0.4 N \div 0.65 N$) 
one finds typical liquid currencies. 
Finally, the smallest values of $\lambda_N$ correspond to the USD and other 
currencies that are strongly tied to US dollar. 
It should be stressed that similar behavior of eigenvalues has been
found for smaller subsectors of the analyzed currencies. In particular,
we have also calculated spectra of tradable and non-tradable (type $C$)
subsectors.
Finally, the second largest eigenvalues ($\lambda_{N-1}$) were
calculated and they are not so clearly separated from the whole spectrum
in two cases:
({\em i}) for currencies with very large $\lambda_N$ {\em i.e.}
with strong independent drift, like GHC and
({\em ii}) for USD and currencies closely tied to it.

In addition to real currencies we have calculated two artificial cases.
The first one, denoted as {\em r.m.} is for all the shuffled time series,
with all time correlations destroyed. As a result we obtained CM
spectrum almost identical as from the random matrix theory.
In the second case we have assumed the Gaussian noise as the exchange rate
of a fictitious currency ({\em fict}) to the USD. The exchange rates to
all other currencies were computed via USD.
Clearly, in this way the original time correlations of USD were modified.
But due to homogeneity of Gaussian noise non-trivial time correlations
did survive. In effect, the CM spectrum for our fictitious currency is
quite similar to typical real currencies. In terms of Fig.~1 the maximal
eigenvalue for this currency is close to the borderline between 
liquid and drifting currencies.


The Minimal Spanning Tree graphs were introduced in graph theory quite long
ago \cite{Kruskal1956}.
Later they were rediscovered several times \cite{Papadimitru,West96}.
To analyze the stock market correlations they were applied by
Mantegna \cite{MantegnaMST} and just recently for FX correlations
\cite{APP2006,McDonald2005}. However, in \cite{McDonald2005}, due to
small number of currencies, all graphs were connected in one bigger
graph.

To construct the MST graph we choose the following metric
\begin{equation}
\label{MSTmetric}
d^X(A,B) = \sqrt{ (1 - C_{AB}^X)/2 } \ , \quad
0 \le d^X(A,B) \le 1 \ ,
\end{equation}
where $X$ denotes the base currency. 
The distance between two time series is smaller if their correlation
coefficient is closer to unity. To each time series corresponds one node
(vertex) in the graph.
We connect two nodes, $A$ and $B$, with a line (``leg'') if their distance
$d^X(A,B)$ is the smallest. In the next step we look for another two
closest nodes and again we connect them with a line. This procedure is
repeated until we obtain a connected tree graph.
The number of legs attached to a given node we call {\em multiplicity of that node},
that will be denoted by $K$.
Clearly, multiplicity of a node is an integer number, $K<N$, where $N$
is total number of nodes in the MST graph.
Sample MST graphs for FX time series were presented in \cite{APP2006}.

Figs.~2-4 display MST graphs for gold, silver and platinum, the oldest traditional 
currencies. 
In all MST graphs for precious metals there is a high multiplicity USD node 
($K\ge 10$). This cluster is extended by the neighboring HKD cluster of considerable
size ($K \ge 7$). 
On the other end of the graph there is more modest EUR node 
(multiplicity $K = 4\div 6$) with European currencies. 
It is interesting to notice that while the currencies CZK, SKK and HUF
have stable position in the euro node, the Polish currency has not. 
In case of gold (XAG) taken as the base currency, PLN is connected to Moroccan dirham (MAD) 
and for silver (XAG) it is connected to Singapore dollar (SGD). 
This indicates a more diversified role of PLN in the world financial system.

\begin{figure}
\begin{center}
  \includegraphics[width=8.5cm,angle=0]{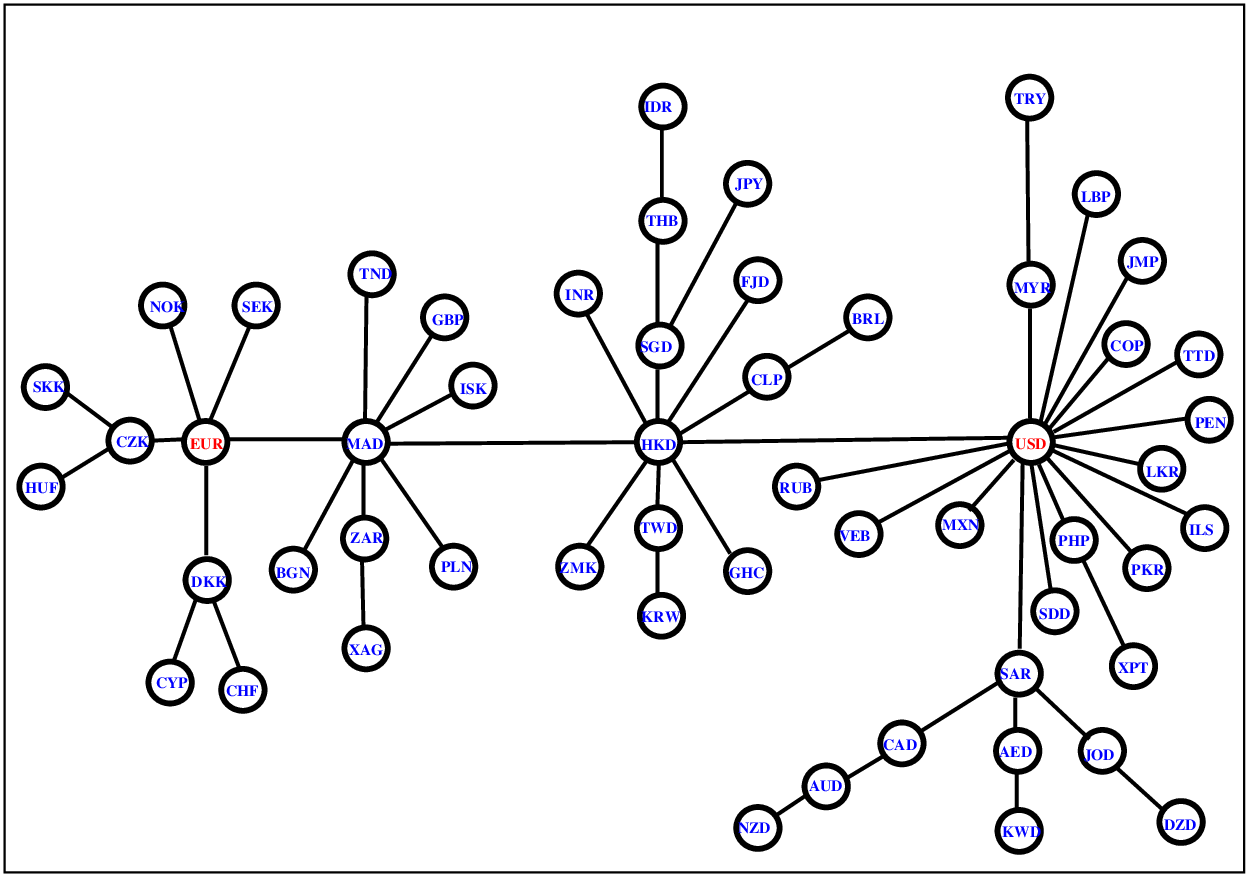}
  \caption{MST graph for XAU taken as the base currency.}
\end{center}
\label{fig:fig2}
\end{figure}

\begin{figure}
\begin{center}
  \includegraphics[width=8.5cm,angle=0]{fig3.eps}
  \caption{MST graph for XAG taken as the base currency.}
\end{center}
\label{fig:fig3}
\end{figure}

\begin{figure}
\begin{center}
  \includegraphics[width=8.5cm,angle=0]{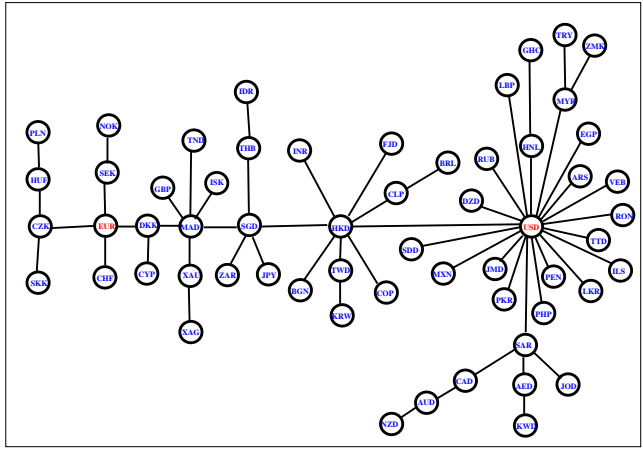}
  \caption{MST graph for XPT taken as the base currency.}
\end{center}
\label{fig:fig4}
\end{figure}

\begin{figure}
\begin{center}
  \includegraphics[width=8.5cm,angle=0]{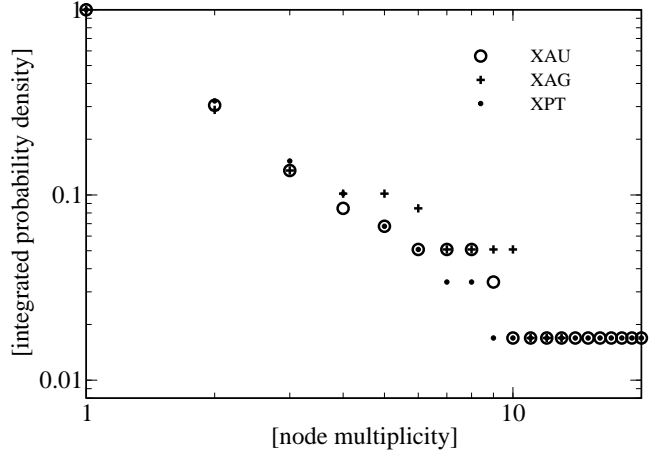}
  \caption{Nodes multiplicity cumulative distribution for gold (XAU), silver (XAG) and platinum (XPT) taken as the base currency.}
\end{center}
\label{fig:fig5}
\end{figure}

\begin{figure}
\begin{center}
  \includegraphics[width=8.5cm,angle=0]{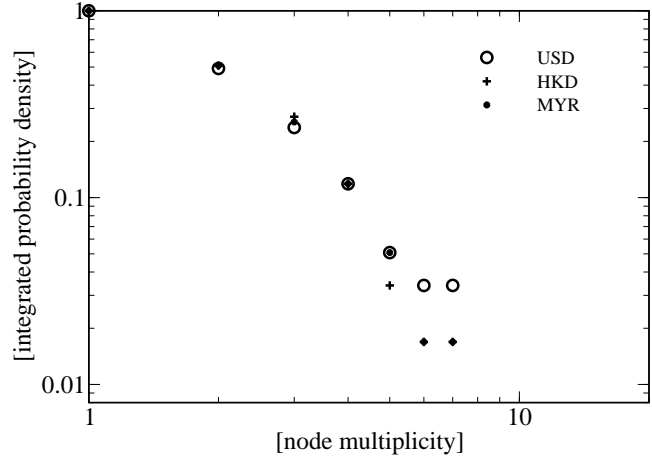}
  \caption{Nodes multiplicity cumulative distribution for currencies with the worst power fits (USD, HKD and MYR).}
\end{center}
\label{fig:fig6}
\end{figure}

\begin{figure}
\begin{center}
  \includegraphics[width=8.5cm,angle=0]{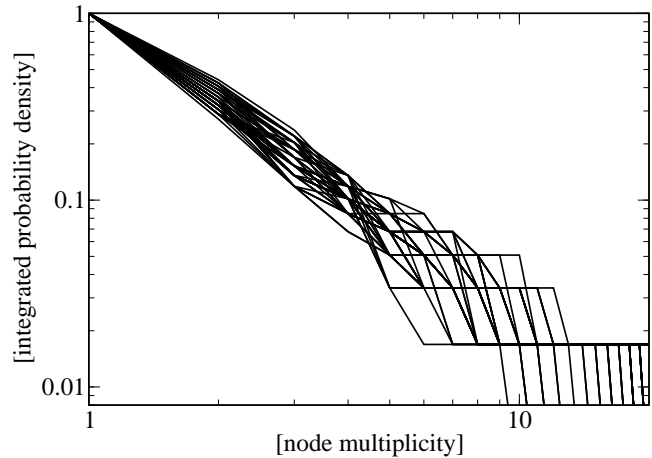}
  \caption{Nodes multiplicity cumulative distribution for remaining 54 currencies taken as the base currency.}
\end{center}
\label{fig:fig7}
\end{figure}

 In our case we have $N = 59$ nodes for a given base currency.
Their multiplicity is denoted by $K_A$. By integer function $N'(K)$
we denote number of nodes with exactly $K$ legs.
Because the total number of nodes, $N$, is relatively small, we introduce
the integrated quantity
\begin{equation}
\label{integratedN}
N(K) = \sum^{K_{max}}_{i=K} N'(K) \ ,
\end{equation}
where $K_{max}$ denotes the maximal number of legs in the MST graph.
$N(K)$ is the number of nodes with $K$ or more legs (a cumulative
distribution with respect to $N'(K)$). Clearly, $N(1) = N$.

To be able to compare graphs with different number of nodes it is
convenient to introduce a normalized (and discrete)
function
$F(K) = N(K) / N$.
For any MST graph we have $F(1) = 1$.
Counting legs in all nodes of a graph we can construct discrete
functions $N(K)$ and $F(K)$.

Now, we are ready to investigate the functional form of the integer
valued discrete function $N(K)$. We have found that on average it
reveals a scale free, inverse power behavior.
However, to estimate quality of the fit we must take into account that
the function has integer values and the inverse power fit by construction
cannot be perfect. 
The smallest possible deviation from a nearest integer value, $\delta(K)$, is
always in the range $0 \le \delta(K) \le 1/2$.
Hence, for the best fit we have the average deviation
$\delta = \langle \delta(K) \rangle_K = 1/4$.
And for the corresponding normalized function
\begin{equation}
\label{DeltaF}
\Delta F = \langle F(K) \rangle_K = \frac{\delta}{N} \simeq 0.004\ .
\end{equation}
Because by definition, $F(K) \le 1$ eq. (\ref{DeltaF}) implies that the
relative error, in the statistical (large $N$) limit should be: $\Delta F/F>0.4$\%.
This result means that even in the case of very large number of legs one cannot obtain better
fit than about half percent.
In our case we have only $59$ legs, a number far too small to have more than a few nodes with higher
multiplicity. By inspection of graphs in Figs.~2-4 one can see only single nodes
with multiplicities $\ge 10$.
Hence, we are very far from what could be called the "large $N$ limit" for the MST graphs.
As a result, even for the best fits one should expect that its accuracy must be much worse
than in the ideal case of half percent.
Especially for the inverse power like fits, where the tail is relatively long and significant.
In this case one can reasonably expect discrepancies around $5$\% for a good fit.
\begin{table}
\begin{center}
  \caption{\label{tab:table1} Power fits for currencies of different
  groups: $A^*, A, B$ and $C$.}
      \begin{tabular}{|lcccc|}%
      \hline %
      \hline %
      \ base currency & $\alpha$ & $\Delta\alpha$ & $\Delta\alpha/\alpha$ & $\lambda_N$\\
      \hline%
      \hline%
      \  metals        & 1.42 & 0.09 & 6.5\% & 43.7 \\
      \hline
     \  $A^\star$  & 1.43 & 0.08 & 5.4\% & 32.1 \\
     \ \ \ $A$        & 1.39 & 0.11 & 7.7\% & 23.2 \\
     \ \ \ $B$        & 1.34 & 0.08 & 6.0\% & 29.9 \\
     \ \ \ $C$        & 1.33 & 0.08 & 5.9\% & 27.4 \\
     \hline
     \ \ \ average    & 1.43 & 0.09 & 6.3\% & 27.3 \\
     \hline
     \ r. m.          & 2.33 & 0.63 & 27\%  & 1.4  \\
     \ fict.          & 1.55 & 0.08 & 5.4\% & 39.0 \\
     \hline
     \hline
\end{tabular}
\end{center}
\end{table}

Inverse power like fits for metals (XAU, XAG and XPT) and for the worst fits
(USD, HKD and MYR) are shown in Figs.~5 and 6. In the latter case the tail
(nodes of multiplicity $\ge 10$) is absent and this is the reason for bad fit.
The currencies HKD and MYR are strongly connected to USD.
In Fig.~7  the data for all 60 base currencies are plotted. With a few exceptions
there is good power fit for nodes' multiplicity distribution.
Detailed numerical data, including the average power exponent
$\alpha$, its standard error ($\Delta\alpha$), and average values of the maximal
CM eigenvalues for all four groups of currencies ($A^\star$, $A$, $B$ and $C$)
are given in Table~1. The average value for $\alpha$ is equal $1.43$, and the
relative errors are only slightly higher than suggested $5$\%.
For the worst fits (USD, HKD and MYR) the corresponding relative errors were found
about $10$\% (USD) and higher (for HKD and MYR).
In addition, results for the shuffled time series and thus destroyed correlations (case {\em r.m.})
and with the base currency being the Gaussian noise are given at the bottom of Table~1.
In the former case, the error was about twice bigger as the worst fit for the real
currencies -- there is no scaling.


 To summarize, it has been found that power like fit and scale free behavior
is fairly good for majority of base currencies.
The worst scaling, $\Delta\alpha/\alpha \ge 9$\%, was obtained for USD
and currencies tied to it (HKD, MYR, JMD, LBP, MAD).
However, even in the worst case ($14$\%) this is much lower value than
for the case of shuffled time series (r.m.), where the relative
error is about $27$\%.
The reason is, that taking USD as the base currency one eliminates its
node from the MST graph. Hence, one of largest clusters (i.e. large
multiplicity node) is missing and the tail of the distribution is
getting thiner.
Because of the relatively modest total number of nodes this effect
is quite significant.
The effect is more drastic for shuffled time correlations, where there
is no power like tail at all.
On the other hand, strong clustering of a currency means that it
is influential for the FX market.
Taking into account that the total number of nodes (number of
currencies) cannot be very big, the standard error of order of 
a few percent should be considered as fairly low.

Similar power like scaling has been obtained by other authors
for other complex networks, with the scaling exponents $\alpha$ 
in the range between $1.1$ and $1.7$ \cite{alphaExmpl1,alphaExmpl2}.
In our calculations the corresponding range is 
$1.3 < \alpha < 2.3$. However, for the two base currencies with $\alpha \ge 2$
(HKD and MYR) the power fit is not so good.

\end{document}